\documentclass[review, 10pt]{elsarticle}

\usepackage{times,amsmath,epsfig}
\usepackage{amssymb}
\usepackage{graphicx}
\usepackage{subfigure}
\usepackage{amsmath}
\usepackage{amsfonts}

\usepackage{algorithmic}
\usepackage{algorithm}
\usepackage{color}

\newtheorem{mytheorem}{\hspace{-11pt}\bf Theorem}

\newtheorem{mydefinition}{\hspace{-11pt}\bf Definition}
\newproof{pf}{Proof}
\newcommand{\EE}{{\mathbb E}}
\newcommand{\calS}{\mathcal{S}}
\newcommand{\calD}{\mathcal{D}}
\newcommand{\bfMu}{\boldsymbol{\mu}}

\journal{Ad Hoc Networks}
\begin{document}

\begin{frontmatter}



\title{Asymptotic Analysis of Cooperative Censoring Policies in Sensor Networks}


\author[uc3m]{Jesus~Fernandez-Bes\corref{cor1}}
\ead{jesusfbes@tsc.uc3m.es}

\author[tuidelft]{Roc\'{\i}o~Arroyo-Valles}
\ead{m.d.r.arroyovalles@tudelft.nl}

\author[uc3m]{Jes\'us~Cid-Sueiro}
\ead{jcid@tsc.uc3m.es}

\address[uc3m]{Signal Theory and Communications Department, Universidad Carlos III de Madrid, Avda. de la Universidad 30, 28911, Legan\'es, Madrid, Spain}
\address[tuidelft]{Faculty of Electrical Engineering, Mathematics and Computer Science, Delft University of Technology, Mekelweg 4, 2628 CD Delft, The Netherlands}

\cortext[cor1]{Corresponding Author: Tel.: +34 91 624 6005; fax:+34 91 624 8749.}

\begin{abstract}
The problem of cooperative data censoring in battery-powered multihop sensor networks is analyzed in this paper. We are interested in scenarios where nodes generate messages (which are related to the sensor measurements) that can be graded with some importance value. Less important messages can be censored in order to save energy for later communications. The problem is modeled using a joint Markov Decision Process of the whole network dynamics, and a theoretically optimal censoring policy, which maximizes a long-term reward, is found. Though the optimal censoring rules are computationally prohibitive, our analysis suggests that, under some conditions, they can be approximated by a finite collection of constant-threshold rules. A centralized algorithm for the computation of these thresholds is proposed. The experimental simulations show that cooperative censoring policies are energy-efficient, and outperform other non-cooperative schemes.

\end{abstract}


\end{frontmatter}


\section{Introduction}
\label{sec:introduction}

Energy stands out as the most critical issue in Wireless Sensor Networks (WSNs) because sensors are usually battery-powered, and consequently, energy consumption limits their operational lifetime. Given the high energy cost of radio transmissions \cite{Raghavendra04}, a great number of {methods} to minimize the energy expenditure due to communication processes has been proposed in the literature (e.g.  \cite{Michelusi12, Orhan12, Zafer09, Liang10, Tutuncuoglu12}).

Among all the state-of-the-art techniques, selective communication {strategies}, also known in the literature as \emph{censoring} {policies} in the context of detection and cognitive radio networks {(e.g., \cite{Appadwedula2005,msechu2012sensor, Maleki11})}, are a promising energy-saving approach. Such policies assume that nodes are able to quantify the relevance (importance) of the incoming messages and discard the low-importance ones to save energy, with the expectation of transmitting more relevant upcoming messages {later}. This importance value can be, for instance, the traffic priority of the routing protocol, the deviation from the mean in a distributed estimation scenario \cite{imer2010optimal}, or the likelihood ratio in a decentralized detection scenario \cite{Appadwedula2005}.


These decisions about transmitting or censoring a message change the amount of energy stored in the batteries and, therefore, have an impact not only on the current {state of the sensor battery} but also on future ones. Intuitively, this implies that current decisions have to be made not only taking into account the instantaneous cost/reward, but also the (expected) impact on future costs/rewards. Mathematically, this {implies} that the sequential decision problem has to be handled using Dynamic Programming (DP) tools. In fact, the DP nature of the problem is also present in several works that deal with censoring communications strategies in WSNs. For instance, \cite{Shrestha11} uses a Markov Decision Process (MDP) {formulation} to design a congestion-aware medium access protocol. \cite{Phan10} exploits an MDP formulation to design transmission policies and adjusts the backoff according to the channel conditions. In the context of replenishable sensor networks, in \cite{Lei09} a Markov chain model to categorize the energy transition is used to derive an optimal single-hop transmission policy. {Some other works also formulate energy-efficient communication policies by means of DP tools. The authors of \cite{Chen07} propose a transmission scheduling algorithm as a stochastic shortest path MDP. Another example is the energy-efficient MDP-based scheduling policies developed in \cite{Li11}, which choose the appropriate transmission mode in energy-harvesting scenarios.}


Most of the censoring {schemes currently available in the literature have been specifically designed for} single-hop networks. The optimization of transmission policies in multihop networks is a more difficult problem because it requires the coordinated decision of all nodes involved in the communications from a source node to its destination. {Censoring rules that optimize point-to-point communications can be shown to be suboptimal from the point of view of the global network performance (i.e., a collection of local optimizers is not a global optimizer).}

{Nevertheless, few algorithms that work in multihop scenarios have been proposed \cite{ArroyoVallesEtAl09, ArroyoVallesEtAl11}}. {The main idea of these works consists of the simply use of local feedback coming from}  {the decisions made by neighboring nodes, so that censoring policies for single-hop networks can be adapted to be more efficient in multihop networks: the policy used by a single node influences the transmission policies of other nodes and produces an overall beneficial coordination effect. Despite of this, the behavior of each node in these schemes is \emph{selfish} in the sense that each node optimizes its own performance measure, which is based on the messages it only processes.}

{A very preliminary approach to the cooperative censoring problem has been proposed in \cite{Fernandez-Bes11}.} {In this approach, the analysis is restricted to very simple line networks, where it is assumed that the nodes closest to the sink will be the first ones that deplete batteries, and all nodes are assumed to use a common censoring threshold. Despite its simplicity,} the experimental results show the potential benefit of exploiting node cooperation in order to save energy and optimize the global performance in multihop networks.

{The main goal of this paper is to analyze the cooperative censoring problem in general multihop networks. In relation to the seminal work in \cite{Fernandez-Bes11}, we carry out three major generalizations: (1) we do not make any assumption about {the battery depletion order in the network, and any arbitrary node depletion order is} allowed, (2) we remove any constraint on the threshold values, and, more importantly, (3) we extend the analysis to more general network topologies, with a special focus on tree networks.}

Both the theoretical and experimental analysis suggest that the optimal censoring policies can be approximated by a collection of piecewise-constant threshold functions of the energy distribution. Unfortunately, the value of these thresholds depends on the expected evolution of the network topology during its whole lifetime. This finding makes the computation of the optimal thresholds difficult. For this reason, we propose a centralized threshold estimation algorithm that is particularized for a tree topology.


From a practical point of view, the proposed algorithm has some major limitations: (1) it works in a centralized manner, (2) it requires some knowledge about the network topology, and it depends on energy and data statistics that may be unknown during operation, and (3) it is computationally complex. Our main contribution in this paper is to show the potential advantages of cooperative censoring schemes, to understand the behavior underlying optimal cooperation, and to promote further research on decentralized adaptive schemes that could work on arbitrary topologies.

The rest of the paper is organized as follows: Section \ref{S:Sensor_model} describes the {network} model for WSNs using an MDP framework. The optimization problem under stationarity conditions is solved in Section \ref{S:Opt_sel_fw} and the analysis of an illustrative scenario {is presented}. For large battery scenarios, an algorithm for computing the asymptotic thresholds, which is based on the stationary node lifetimes, is proposed in Section \ref{S:ApproximateCensoringPolicy}. Simulation results for different network configurations are presented in Section \ref{S:Simulations}, and Section \ref{S:Conclusions} wraps up the paper including some pointers {for} future work.

\section{{ Network model}}\label{S:Sensor_model}
For the purpose of the analysis that follows, we consider a sensor network as a collection of sensor nodes $\mathcal{N}=\{i|i=1,\ldots,N\}$ and a set of edges, $\mathcal{E}\subset \mathcal{N}\times \mathcal{N}$. We assume that the network operation time is divided into epochs, {which are} denoted by counter variable $k$. At each epoch, at most one message is generated by just one of the nodes. Messages are generated at different nodes, which are able to make decisions about routing (or not routing) these messages to their destination. The duration of each time epoch is assumed to be long enough to ensure that the message arrives to its destination {generally after some hops}. Some results of the treatment of simultaneous transmissions in censoring communications under more realistic scenarios can be found in \cite{hansen2010testing}. 


As explained in the introduction, since we are interested in maximizing a long-term reward and current actions have an impact on future states, our problem will fall into the DP framework. Moreover, given that the state dynamics are assumed Markovian, the problem will be modeled as an MDP. In the following sections we present all the components of the {MDP: state space, action space, state dynamics and rewards.}

{To facilitate the readability of this paper, the most relevant notation introduced here and in the following sections is summarized in Table \ref{Table_Notation}.}

\subsection{Network state}


The network state at epoch $k$ will be characterized by three main variables:

\begin{itemize}
 \item ${\bf e}_k$: {$N$-dimensional vector} of the available energies (battery levels) at epoch $k$, where {$e_{k,i}$} denotes the available energy at node $i$. It reflects the ``internal state'' of {each} node. {We assume that all energy values are multiple of a basic energy unit, $\epsilon$. For simplicity, we will take $\epsilon=1$, so that ${\bf e}_k$ is a vector of integers}.
 \item $y_k \in \mathcal{N}$: the index of the message source. If node $j$ is the source of the current message, then $y_k=j$. The notation $y_k=0$ will be used for the event {in which} no message has been generated by any node at epoch $k$.
 \item $x_k$: importance assigned to a message {which is generated} at epoch $k$. For mathematical convenience, we assume $x_k=0$ when $y_k=0$. {The importance value is application-dependent (see \cite{ArroyoVallesEtAl11} for some examples).}
\end{itemize}

Besides ${\bf e}_k$, $y_k$ and $x_k$, {any other additional information about the message which could be relevant to estimate the energy cost of transmitting it, such as the packet length, may be used for the network to make decisions.} This additional information, together with $x_k$ and $y_k$, is collected into vector ${\bf z}_k$. Following the usual terminology in MDP models, the network state vector is defined as ${\bf s}_k = ({\bf e}_k,{\bf z}_k)$; i.e., the state vector contains all and only the information that is {available to} make a decision at epoch $k$. The set of all possible states is denoted as {${\mathcal T}$}.


\subsection{Actions and policies}

At epoch $k$, the network must make a decision $a_k$ about sending or not {sending} the current message. The message is sent if $a_k=1$,  while it is censored (and definitely dropped) if~$a_k=0$. A {transmission} policy $\pi = \{a_{1},a_{2},\ldots \}$ at a given node is a sequence of decision rules, which are functions of the state vector; i.e.,
\begin{equation}
a_k = \pi_k({\bf s}_k) = \pi_k({\bf e}_k,{\bf z}_k).
\label{Eg.General0Trasmitter}
\end{equation}

Also, note that decisions are made at the network level. If a message which is generated at node $i$ is transmitted to a neighbor, it will be forwarded towards its destination. Although one could think of providing nodes with the capability of censoring messages received from other nodes, from the point of view of the overall network performance, this {behavior} is always suboptimal. If node $j$ censors a message from node $i$, node $i$ should have previously censored this message and saved the corresponding energy.

\subsection{State dynamics}

We are specially interested in the energy dynamics given the decisions. At each time epoch, nodes consume an amount of energy which depends on the taken actions. We will express the available energy at epoch $k$ recursively as
\begin{equation}
{\bf e}_{k+1} = {
                ({\bf e}_k - a_k {\bf c}_{1k} - (1-a_k) {\bf c}_{0k})^+},
\label{Eq:EnergyDynamics}
\end{equation}
where $(z)^+ =\max\{z,0\}$, for any $z$, ${\bf c}_{1k}$ contains the energy consumed by all the nodes when the network decides to transmit the message, and ${\bf c}_{0k}$ is the energy consumed when the message is censored. Both, ${\bf c}_{0k}$ and ${\bf c}_{1k}$, are vectors of integer components. Vector ${\bf c}_{0k}$ includes the cost of data sensing at the source node, {any} \emph{background} consumption due to other activities in the node, and any other cost incurred during the last epoch. Parameter ${\bf c}_{1k}$ accounts for all the previous costs plus the cost of {routing the message from the source node to its destination (so that ${\bf c}_{1k}\ge {\bf c}_{0k}$ in a component-wise comparison). This may include the transmission costs of all the nodes in the path to the destination, the subsequent reception costs, and any other collateral cost at the nodes that does not belong to the transmission path (for instance, additional energy consumption due to overhearing messages intended for other nodes). In general, we assume that ${\bf c}_{1k}$ and ${\bf c}_{0k}$ are stochastic stationary and {i.i.d. processes, though they can} depend on ${\bf z}_k$.

We assume that, when $y_k=0$, ${\bf c}_{0k}={\bf c}_{1k}$ because there are no true message transmissions associated to $a_k=1$ in this case. Note that, in general, nodes may consume some energy even though no message is generated in the network.


With regard to the other component of the state vector, i.e., ${\bf z}_k$; we assume that it is a stationary {i.i.d. process}, and independent of ${\bf e}_{k-\ell}$ or ${a}_{k-\ell}$, for any $\ell>0$.

{All this information, along with the probability functions {of ${\bf c}_{0k}$, ${\bf c}_{1k}$ and ${\bf z}_k$, which are denoted by $P_0$, $P_1$, and $p$, respectively,} are enough to characterize the state transition probabilities, $p({\bf s}_{k+1}| {\bf s}_k, {a_k})$. For instance, if {$e_{k,i} > e_{k+1,i} \ge 0$ $\forall i$},  we have}
\begin{align}
p({\bf s}_{k+1}| &{\bf s}_k, {a_k})
     = ({a_k P_1}({\bf e}_k-{\bf e}_{k+1}|{\bf z}_{k}) 
    + (1-{a_k) P_0}({\bf e}_k-{\bf e}_{k+1}|{\bf z}_{k}))
       p({\bf z}_{k+1}). 
\label{Eq:StateDynamics}
\end{align}


Though for the theoretical analysis we assume that the distributions {$P_0$, $P_1$ and $p$ are known,} in practice only some statistics, which may be estimated from the data, {are} required.

\subsection{Rewards}
\label{S:Rewards}

{
The {\em reward} at time $k$ is given by
\begin{equation}
r_k = x_k q_k {a_k},
\label{EqReward}
\end{equation}
where $q_k \in \{0,1\}$ denotes the {\em success index} (a binary variable {whose value is} 1 if the transmission is successful, and zero otherwise). The transmission is considered successful whenever a message arrives to its destination. On the contrary, $q_k=0$ if, for any reason, the message is lost: imperfect communication channels, battery depletion of a relay node in the transmission path, etc.

 
The figure of merit to design the transmission policy will be the importance sum of all messages {\em successfully} delivered to its destination. Accordingly, the total reward up to epoch $k$ is defined as
\begin{equation}
R_k = \sum_{\ell=0}^{k} r_{\ell} 
    = \sum_{\ell=0}^{k} {a}_{\ell}  q_{\ell}  x_{\ell} ,
\label{EqSk}
\end{equation}
and, for any initial state ${\bf s}_{0}=({\bf e}_{0},{x}_{0})$ the transmission policy $\pi$ is chosen in order to maximize the total expected reward, defined as
\begin{equation}
V_\pi{({\bf s}_{0}) = \EE\left\{R_{\infty}|{\bf s}_{0}\right\} = \EE\left\{\lim_{k\rightarrow \infty} R_k|{\bf s}_{0}\right\}}.
\label{EqSOrder2}
\end{equation}

Note that, since nodes have limited energy resources, the sum in (\ref{EqSk}) only contains a finite number of nonzero values (eventually, for some $k$, {${\bf e}_k={\bf 0}$}, and $\forall k'\geq k$, we have $r_{k'}=0$).

Note also that, making a difference with other MDP models, the total reward does not include a discount factor and, thus, future rewards are not penalized. This is a reasonable assumption whenever the expected network lifetime is moderate. For very low values of {${\bf c}_0$}, i.e., as the energy cost of censoring becomes arbitrarily close to zero, it can be shown that the optimal censoring rule will transmit only messages with extremely high importance, and will censor almost every message. In such situations, a penalized discount should be used instead of {\eqref{EqSk}}. In practice, since the energy costs are usually not negligible even under a censoring decision {\cite{raghunathan2002energy}}, no degeneracy problems arise from using \eqref{EqSk}}.

\section{Optimal transmission policies}\label{S:Opt_sel_fw}

\subsection{Markov Decision Process}

As stated, the tuple defined by $({\mathcal T},{\mathcal A},P,r)$, where ${\mathcal T}$ is the set of states, ${\mathcal A}=\{0,1\}$ is the set of possible decisions (actions), $P$ is the transition probability measure given by (\ref{Eq:StateDynamics}) and $r$ is the reward function, has the structure of an MDP. Moreover, since the action set ${\mathcal A}$ is finite, an optimal policy exists and it is Markovian. This means that there is an optimal policy such that, at any epoch $k$, the decision rule only depends on the state ${\bf s}_k$ \cite{Puterman:05}. 

{
The optimal decision policy $\pi^*$ is such that, for any state ${\bf s}$, it satisfies the Bellman's optimality equation }
{
\begin{eqnarray}
V_{\pi^{*}}({\bf s}) 
   = \max_{{a} \in \{0,1\}} 
           \left\{ \EE\left\lbrace 
                        r_k + V_{\pi^{*}}({{\bf s}_{k+1}) |a_k=a, {\bf s}_k}={\bf s}
                   \right\rbrace  \right\}.
\label{Eq.Bellman}
\end{eqnarray}
}
{
The solution of this optimality equation is presented in the following theorem. }

\begin{mytheorem} \label{Th.General}
{\em {Let $\{\mathbf{z}_k,k\ge 0\}$ $\{{\bf c}_{0k},k\ge~0~\}$ and $\{{\bf c}_{1k},k\ge 0\}$ be jointly stationary and time-independent processes, and let ${\bf e}_k$ be the energy process given by (\ref{Eq:EnergyDynamics}). {Further}, for any state ${\bf s} = ({\bf e},{\bf z})$, {let} assume that $Q({\bf s}) = \EE\{q_k | {\bf s}_k={\bf s}\}$ {(i.e., the probability of success)} does not depend on $k$.}

Consider the stationary policy given by
\begin{equation}
\pi^*({\bf s}) = u(Q({\bf s}) x - \mu({\bf s})),
\label{Eq.opt.dk}
\end{equation}
where $x$ is the importance value contained in ${\bf z}$, {$u$ stands for the Heaviside step function} and threshold $\mu$ is defined recursively through the pair of equations
{\begin{align}
\mu({\bf s}) =& \EE\left\{\lambda\left(({\bf e}-{\bf c}_{0k})^+\right) 
                   - \lambda\left(({\bf e}-{\bf c}_{1k})^+\right)|{\bf s}\right\}
\label{Eq.thresh_theorem}
\\
\lambda({\bf e}) =& \EE\{\lambda\left(({\bf e}-{\bf c}_{0k})^+\right) 
                 +  (Q({\bf s}) x-\mu({\bf s}))^+ |{\bf e}\} ,
\label{Eq.Th1lambda}
\end{align}}
with $(z)^+ = \max\{z,0\}$, for any $z$.

Policy $\pi^*$ maximizes $V_\pi$ (given by (\ref{EqSOrder2})).

The auxiliary function $\lambda({\bf e})$ represents the total reward that can be expected for an initial battery load ${\bf e}$, i.e.,
\begin{eqnarray}
\lambda({\bf e}) = \EE\{V_{\pi^*}({\bf s}_{0}) |{\bf e}_{0}={\bf e} \}.
\label{EqDefLambda}
\end{eqnarray}}
\end{mytheorem}

\begin{pf}
See Appendix \ref{Th.General}. 
\end{pf}

The optimal censoring policy given by \eqref{Eq.opt.dk} compares the importance of the message with a threshold.
Note that functions $\lambda$ and $\mu$ can be computed iteratively through a backward recursion, starting from low energy vectors towards high energy vectors. However, from a practical point of view this is problematic because energy values are initially high. Nevertheless, the observation of these functions in some illustrative cases provides some clues for the design of censoring policies that are computationally more efficient.


\subsection{Analysis of a basic scenario}
\label{Sec:CEC}

In order to understand the characteristics of the optimal decision policy, we analyze the following simplified model in a {time-stationary} network composed of $N$ nodes {(i.e., the time subscript $k$ can be drop off from some variables)}:
\begin{enumerate}
\renewcommand{\theenumi}{A\arabic{enumi}}
\item The energy consumptions are deterministic. Let ${\bf c}_0(j)$ and ${\bf c}_1(j)$ denote {the energy expenditure vector (of size  $N\times 1$)} when a message (generated by source $j$) is rejected and transmitted, respectively. Accordingly, we denote ${\bf c}_0(0)$ ($={\bf c}_1(0)$) as the consumption in the silent epochs, i.e., when no message is generated in the network.
\item All messages emanated from a given node are routed through the same path towards a fixed destination node. Thus, for each source node $j$, we can define a binary vector {$\textbf{v}(j)$ of size $N\times 1$, such that $v_{i}(j)=1$, if messages originated at node $j$ are routed through node $i$, and 0 otherwise.} 
\item If nodes have enough energy to deliver a message to the destination, the probability of success is 1 (i.e., there are no lost messages). Mathematically:
\begin{align}
{Q({\bf e},y_k) = \EE\{q | {\bf e},{\bf z}\} = {\prod_{v_i(y_k)=1}} u(e_i-c_{1,i}(y_k)),}
\end{align}
where $y_k{\in\mathcal{N}}$ corresponds to the index of the source node {at epoch $k$, and $c_{1,i}(y_k)$ is the element $i$ of vector ${\bf c}_1(y_k)$}.
\end{enumerate}

Assumption A2 includes scenarios such as sensor networks with a unique sink or multiple sinks, where all the messages from a given source node are routed to the same destination through the same set of nodes (fixed routes). 

Since $Q({\bf e},y_k)$ is binary, {policy \eqref{Eq.opt.dk}} can be written as
\begin{align}
{a_k} = (x_k - \mu({\bf e},y_k))^+ \cdot Q({\bf e},y_k).
\label{Eq.opt.dk.Stat.Ej}
\end{align}

Also, since energy consumptions are deterministic, \eqref{Eq.thresh_theorem} becomes
{\begin{gather}
\mu({\bf e},y_k) = \lambda(({\bf e}-{\bf c}_0(y_k))^+) - \lambda(({\bf e}-{\bf c}_1(y_k))^+).
\label{Eq.Mu_Stat.Ej}
\end{gather}}

Defining the frequency of message generation by each node as $p_j=P\left(y_k=j\right)$, 
{\begin{align}
\EE\{\lambda(({\bf e}-{\bf c}_{0}(y_k))^+)\} = \sum_{j={0}}^{N} {p_j}\lambda(({\bf e}-{\bf c}_0(j))^+)   \\
\EE\{\lambda(({\bf e}-{\bf c}_{1}(y_k))^+)\} = \sum_{j={0}}^{N} {p_j}\lambda(({\bf e}-{\bf c}_1(j))^+).
\end{align}}
Thus, defining $h_j(\mu({\bf e},j)) = \EE\{(x_k-\mu({\bf e},j))^+ |y_k=j\}$, \eqref{Eq.Th1lambda} can be expressed as
{\begin{align}
\lambda({\bf e})
    &= \sum_{j={0}}^{N} {p_j} \lambda(({\bf e}-{\bf c}_0(j))^+)        
    + \sum_{j={0}}^{N} {p_j} \EE\{(Q({\bf e},j) x_k-\mu({\bf e},j))^+ |y_k=j\}        \nonumber\\
    &= \sum_{j={0}}^{N} {p_j} \lambda(({\bf e}-{\bf c}_0(j))^+)                             
    + \sum_{j={0}}^{N} {p_j} \EE\{(x_k-\mu({\bf e},j))^+ |y_k=j\}\cdot Q({\bf e},j) \nonumber\\
    &= \sum_{j={0}}^{N} {p_j} \left(\lambda(({\bf e}-{\bf c}_0(j))^+) 
                                      + h_j(\mu({\bf e},j)) \cdot Q({\bf e},j) \right).
\label{Eq.Lambda_Stat.Ej}
\end{align}}

We use the convention $x_k=0$ when $y_k=0$ and, thus, $h_0(\mu({\bf e},0))=0$, for any $\mu$.

To get some intuition about the shapes of the optimal thresholds, {we consider a simple 2-node line network (i.e., ${\cal N} = \{1,2\}$) so that} node $1$ is connected to node $2$, and node $2$ is connected to a sink, which in turn is connected to a power line. Only nodes 1 and 2 have limited battery resources. The source node of every message is selected at random with equal probability (i.e., $p_1=p_2=0.5$ and $p_0=0$). Also, the importance distribution of the messages is exponential with mean 1 independently of the source node. Energy consumptions are deterministic, and given by ${\bf c}_0(1) = (3,1)^\top$, ${\bf c}_0(2) = (1,3)^\top$, {${\bf c}_1(1) = (11,10)^\top$, ${\bf c}_1(2) = {(1,10)^\top}$}. {Fig. \ref{fig:MuExp} illustrates functions {$\mu_1({\bf e})=\mu({\bf e},1)$} and $\mu_2({\bf e})=\mu({\bf e},2)$, i.e., the threshold function for messages generated at node 1 and 2, respectively.}

\begin{figure}[t]
\centering
\subfigure[]{\includegraphics[trim=10mm 65mm 10mm 60mm, clip,width=1\columnwidth]{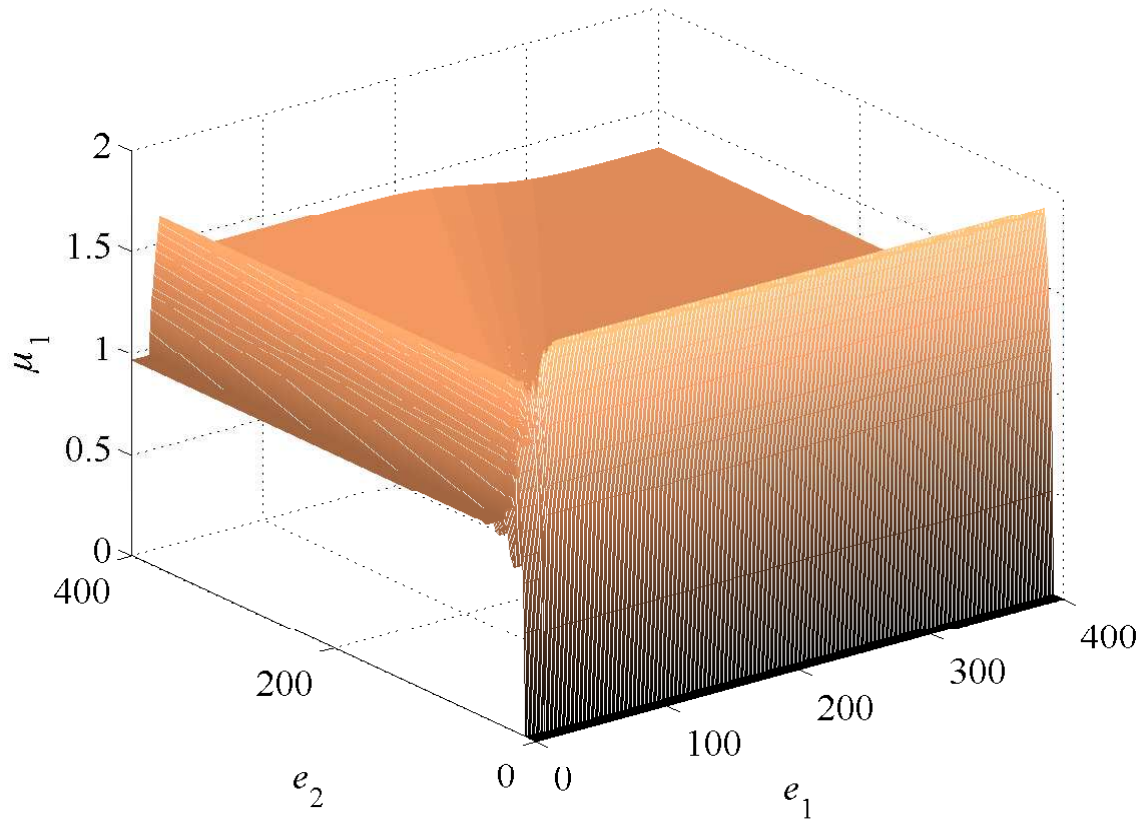}}
\subfigure[]{\includegraphics[trim=10mm 65mm 10mm 60mm, clip,width=1\columnwidth]{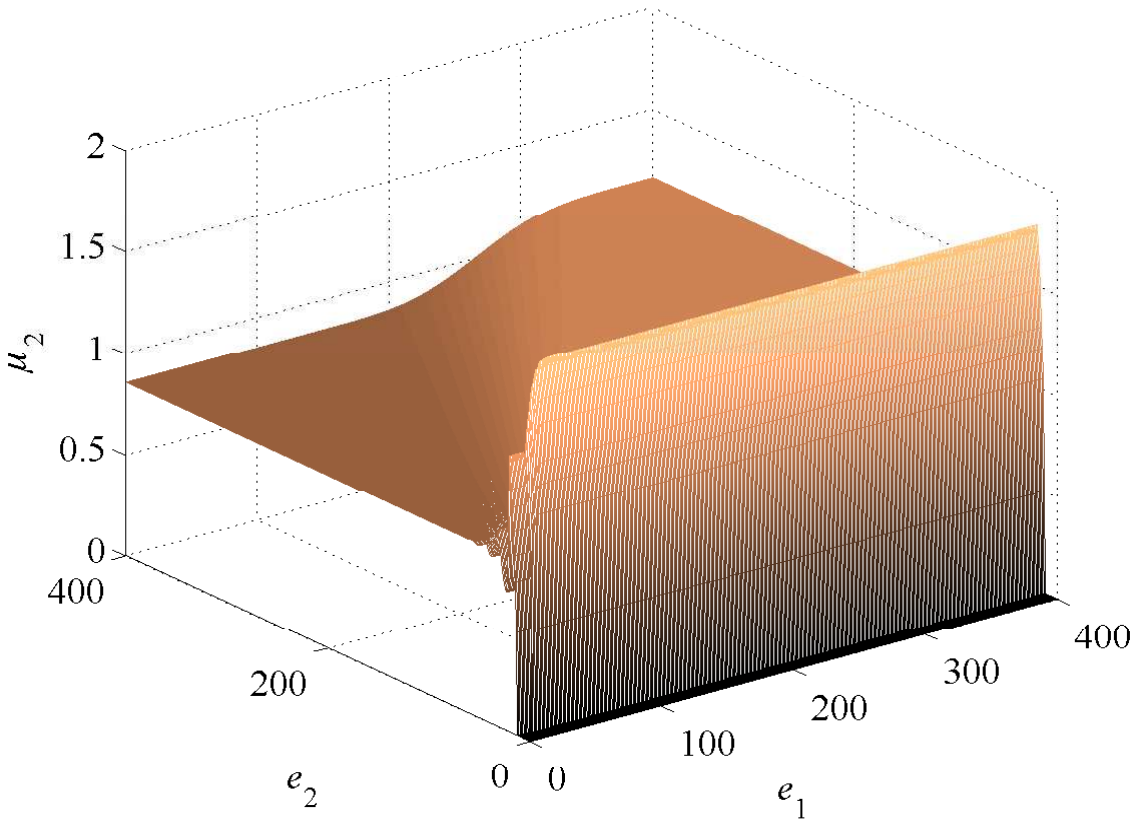}}
\caption[Threshold functions]{Threshold functions for a network of 2 nodes (plus the sink) in a line topology. (a) Optimal threshold function at node 1, {$\mu_1({\bf e})=\mu({\bf e},1)$} for a non-rechargeable node with ${\bf c}_0(1) = (3,1)^\top$, ${\bf c}_0(2) = (1,3)^\top$, ${\bf c}_1(1) = (11,10)^\top$, {${\bf c}_1(2) = (1,10)^\top$}. (b) Optimal threshold function at node 2, {$\mu_2({\bf e})=\mu({\bf e},2)$}.}
\label{fig:MuExp}
\end{figure}

The figure shows two remarkable characteristics:
\begin{enumerate}
\item For small values of $e_1$ or $e_2$ (i.e., {energy values} close to the axes $e_1=0$ or $e_2=0$), the threshold functions show an oscillatory behavior.
\item However, far from these axes, the threshold functions show two large flat regions.
\end{enumerate}
{These characteristics are in agreement with the behavior observed for single nodes in \cite{ArroyoVallesEtAl09}. There, it was shown that the energy level played a key role in the threshold values only when the batteries were close to be exhausted. That is the reason for the oscillatory behavior for low energy values.}

According to \eqref{Eq.Mu_Stat.Ej}, function $\mu$ is a difference of two shifted versions of function $\lambda$. Then, function $\lambda(e)$ is approximately linear whenever $\mu$ is approximately constant. This is observed in Fig. \ref{fig:Lambda}.}
\begin{figure}[t]
\centering
\includegraphics[clip,width=1\columnwidth]{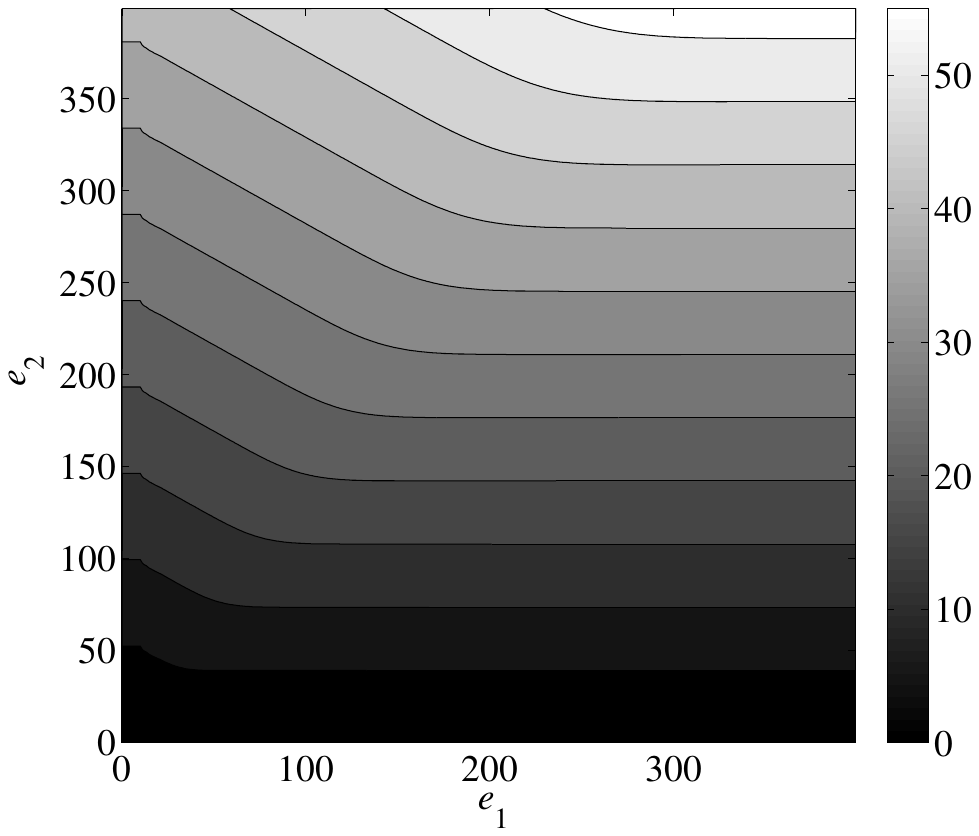}
\caption[Value function]{The value function $\lambda(e)$ for a network of 2 nodes in a line topology for the optimal threshold functions calculated in Fig.\ref{fig:MuExp}.}
\label{fig:Lambda}
\end{figure}

The previous behavior of the threshold functions can be systematically observed for different values of the energy consumption parameters and arbitrary importance distributions, and even for networks with a higher number of nodes. Although the precise shape of the oscillatory behavior near the axes is highly dependent on the importance distribution, these values are only relevant for a few censoring decisions (specifically, those made before the battery depletion). Hence, they have a small influence on the overall performance whenever the initial battery load is high.

This finding suggests a strategy to {approximate the optimal} censoring policy: to approximate the threshold functions by piecewise-constant functions in the energy space. To that aim, we need to solve two problems: (1) characterize each constant-threshold region, and (2) estimate the threshold value in each region.

An intuition about that approximation can be deduced from Figs. \ref{fig:MuExp} and \ref{fig:Lambda} corresponding to the analysis of the basic scenario under study. When $e_1\gg e_2$, it can be expected that node 2 died in the first place (because of battery depletion). If this situation happens, node 1 would get disconnected from the sink, and no more messages would arrive to the destination. On the other hand, if $e_2\gg e_1$, node 1 is expected to be the first to deplete the battery; however, node 2 would keep still connected to the sink afterwards. These different possibilities in the network evolution, which depend on the expected node lifetimes and, more specifically, in the sequence of node battery depletions, make a difference in the value of the threshold functions.

\begin{figure}[t]
\centering
\subfigure[]{\label{fig:LifetimesA} \includegraphics[clip,width=1\columnwidth]{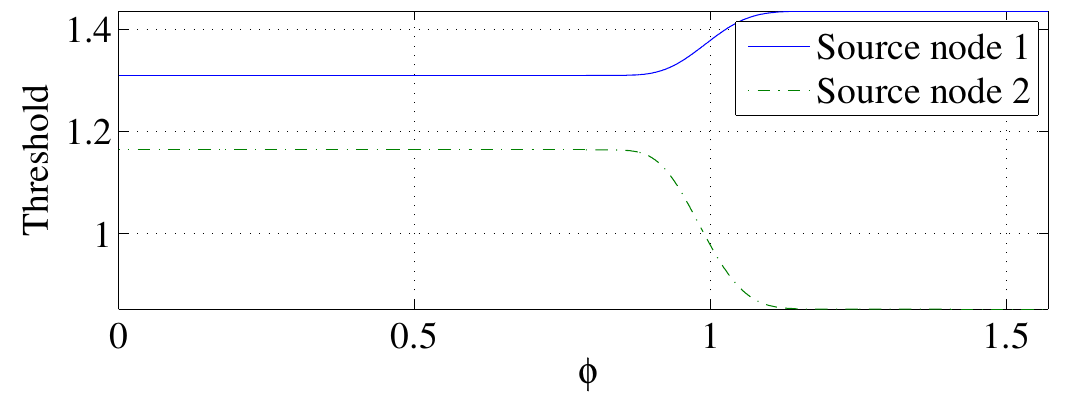}} 
\subfigure[]{\label{fig:LifetimesB} \includegraphics[clip,width=1\columnwidth]{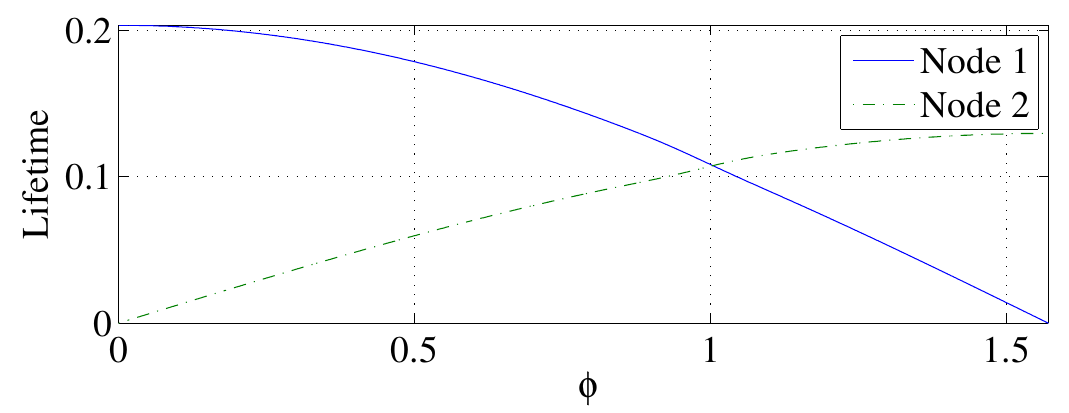}} 
\caption[Lifetimes]{(a) Asymptotic threshold for the 2-node network described in Section \ref{Sec:CEC}. (b) Lifetime of nodes. Direction $\phi$ is {counter-clockwise measured in radians from the origin ${\bf e}=(0,0)$.}}
\label{fig:Lifetimes}
\end{figure}

The comparison of the node lifetimes and the thresholds is also illustrative. Let us express the energy vector in polar coordinates, ${\bf e} = (e_1,e_2)^\top = (r\cos\phi,r\sin\phi)^\top$, where $\phi$ is the direction (counter-clockwise measured in radians from the origin) {and $r$ is the distance to the origin, which is assumed to be large.} Fig. \ref{fig:LifetimesA} shows the value of the asymptotic thresholds (i.e., the threshold value for large energy) as a function of $\phi$ for the 2-node network example discussed in this section. It shows the transition in the threshold function from one constant region to the other one. Fig. \ref{fig:LifetimesB} illustrates the expected lifetime of node 1 and 2 (denoted as $T_1$ and $T_2$, respectively), which is computed assuming that each node uses a constant threshold equal to its corresponding asymptotic value. For the sake of clarity, the details of the computation of these expected lifetimes will be explained in Section \ref{S:FlatRegions}. We can observe that, for $e_1 \gg e_2$, node 1 is expected to live longer than node 2, and vice versa. Further, the transition from $T_1>T_2$ {to $T_2>T_1$} takes place at the same direction as the transition from one constant-threshold region to the other one.

Taking these empirical observations as the starting hypothesis, in the next Section we will develop a threshold estimation algorithm that is based on two main ideas: (1) the energy space can be divided into a set of constant-threshold regions (where $\lambda({\bf e})$ is approximated by a linear function of the energy space), and (2) the threshold value in each region critically depends on which node is expected to be the first to deplete the battery (denoted as the \emph{critical node} in the following).



\section{Approximate censoring policy}
\label{S:ApproximateCensoringPolicy}

To overcome the computational cost of \eqref{Eq.thresh_theorem} and \eqref{Eq.Th1lambda} (or \eqref{Eq.Mu_Stat.Ej} and \eqref{Eq.Lambda_Stat.Ej}), which is exponential in the number of nodes, we propose a censoring policy based on piecewise-constant thresholds which approximate the optimal ones. The main idea is to avoid computing the values of the exact thresholds for low energy values and substitute them by the asymptotic ones in the energy space (i.e., the constant values achieved when the energy level is large enough). 

\subsection{Notation}
\label{SecNotation}

Let us explain the notation that will be used in this section (which is also summarized in Table \ref{Table_Notation}). For any subset of indices, $\mathcal{I},\mathcal{J}\subset \mathcal{N}$ and any arbitrary matrix ${\bf M}$, we define ${\bf M}_\mathcal{J}$ as the submatrix resulting from {taking from ${\bf M}$ the columns corresponding to indices in $\mathcal{J}$}, and ${\bf M}_{\mathcal{IJ}}$ is the result of taking from ${\bf M}$ the rows included in $\mathcal{I}$ and the columns included in $\mathcal{J}$. Also, with some abuse of notation, when the index set is a singleton, $\{i\}$, we will write it simply as $i$. Since vectors can be considered as single-column matrices, a similar notation will be used for vectors.

Also, we define cost matrix ${\bf C}_b$ with columns $\bar{\bf c}_{bj} \doteq \EE\{{\bf c}_{bk}|y_k=j\}$, for~$b\in\{0,1\}, j\in\mathcal{N}$, and
\begin{equation}
{\bf \Delta} = {\bf C}_1^\top - {\bf C}_0^\top,
\end{equation}
which is a $N \times N$ matrix containing the incremental costs of deciding to transmit a message at any node and for any message source.

Finally, we define vector ${\bf p}$, which represents the probability of being a source and with components $p_i = P(y_k=i)$, and threshold vector $\bfMu$, whose components are the threshold values at each node.

\subsection{Assumptions}

From now on, we will ignore assumptions A1-A3 from the previous section, and consider a more general scenario with stochastic energy consumptions in a restricted network topology. 
More specifically, we will employ the following assumptions:
\begin{enumerate}
\renewcommand{\theenumi}{B\arabic{enumi}}
\item Energy consumptions only depend on the source of the message, $\EE\{{\bf c}_{bk}|{\bf z}_k\} = \EE\{{\bf c}_{bk}|y_k\}{<\infty}$ for $b \in \{0,1\}$.
\item For {the minimum value of the energy vector being large enough (i.e., $\min_i\{e_i\}$, $i\in\mathcal{N}$  large)}, $Q({\bf s})$ only depends on the message source. Clearly, if the minimum energy value is too low, the hypothesis is obviously false, as some nodes may not have energy to transmit any message, resulting $Q({\bf s})=0$. We will use the abbreviated notation $Q_y=Q({\bf e},{\bf z})$, where $y\in \mathcal{N}$ is the source of the message. 
\item {A tree-topology network}: all messages generated in the network are forwarded to a sink (root) node through the branches of the tree. {This assumption implies that there is only one destination node to whom all the messages are forwarded through some fixed routes.}
\end{enumerate}
Assumptions B1 and B3 can be relaxed. We will discuss this issue at the end of this Section.

\subsection{Structure of the approximate censoring algorithm}

The structure of the approximate censoring algorithm proposed in this paper is shown in Algorithm \ref{AlgMain}. {The entry-point} function ${\bf main}({\cal G},{\bf e},{\bf p},{\bf C}_0,{\bf C}_1)$, which depends on the network graph, ${\cal G}$, the energy state ${\bf e}$ {of all the nodes} and the MDP model parameters, returns the threshold vector $\bfMu$ and the vector ${\bf w}$ (with the coefficients of a linear approximation of $\lambda({\bf e})$, which will be described in Section \ref{S:linear_approx}). Starting from an initial guess of the threshold vector $\bfMu$, the algorithm proceeds iteratively in two steps: (1) given the thresholds, compute the critical node $i$ and its stationary lifetime $T_i$ (function ${\bf CriticalNode}$, whose pseudocode is shown in Algorithm \ref{AlgLife}), and (2) given $i$ and $T_i$, estimate the corresponding thresholds (function ${\bf Thresholds}$, whose pseudocode is in Algorithm \ref{alg1}). Since the thresholds and the critical node are mutually inter-dependent, the main algorithm should repeat these two steps iteratively until {convergence}.

\begin{algorithm}                      
\caption{Compute 
         $(\bfMu,{\bf w})={\bf main}({\cal G},{\bf e},{\bf p},{\bf C}_0,{\bf C}_1)$}    
\label{AlgMain}                           
\begin{algorithmic}                    
    \STATE $\bfMu = {\bf 0}$
    \REPEAT 
       \STATE $\bfMu_0 = \bfMu$
       
       \COMMENT{Step 1: Identify the critical node and its stationary lifetime given the thresholds}
       \STATE ${(i,T_i) = {\bf CriticalNode}}(\bfMu,  {\cal G},{\bf e},{\bf p},{\bf C}_0,{\bf C}_1)$
       
       \COMMENT{Step 2: Compute the thresholds given the critical node}
       \STATE {$(\bfMu,{\bf w}) = {\bf Thresholds}   ({i, T_i},{\cal G},{\bf e},{\bf p},{\bf C}_0,{\bf C}_1)$} 
     
    \UNTIL $\bfMu=\bfMu_0$
\end{algorithmic}
\end{algorithm}

In the following subsections, we justify and explain these two steps into detail.

\subsection{Characterizing the constant-threshold regions}
\label{S:FlatRegions} 

The empirical analysis in Section \ref{Sec:CEC} has shown that, in order to estimate the constant-threshold regions, the identification of the \emph{critical node} is of major importance. In this section we provide some formal definitions for that purpose.



\begin{mydefinition}
Given an MDP $({\cal T},{\cal A},{\cal P},r)$ for a network with $N$ nodes and an energy vector ${\bf e}$, the lifetime of node $i$ under a stationary policy $\pi$ is the random variable defined as
\begin{equation}
\tau_i({\bf e}) = \sup\left\{K: \sum_{k=0}^{K} g_{k,i} \le e_i \right\},
\end{equation}
where $g_{k,i}$ is the consumption of node $i$ at epoch $k$. 
\end{mydefinition}

We have made the dependency of $\tau_i$ with $e_i$ explicit because, provided that $\EE\{g_{k,i}\}<\infty$, $\tau_i(e_i)$ is a stochastic renewal process (see e.g. \cite{smith1958renewal}). However, it is non-stationary in general because the energy consumption of node $i$ at time $k$ may depend on the activity of other nodes, which may change with time. This makes the computation of the expected node lifetime difficult. Fortunately, for our purposes, a related measure has a straightforward computation: the stationary lifetime.

\begin{mydefinition}
Given an MDP $({\cal T},{\cal A},{\cal P},r)$ for a network with $N$ nodes, the stationary lifetime of node $i$ with energy $e_i$ under a stationary policy $\pi$ is the expected value of the node lifetime, assuming that all the other nodes have infinite energy supply, i.e.,
\begin{equation}
T_i(e_i) = \lim_{\alpha \rightarrow \infty} \EE \{\tau_i(\alpha,\ldots,\alpha,e_i,\alpha,\ldots, \alpha)\}. 
\end{equation}
\end{mydefinition}

Note that the stationary lifetime does not reflect the true expected lifetime of the node because it assumes a stationary situation where all the other nodes have an infinity battery supply. However, if there is a node in the network whose battery is considerably smaller than that of all the other nodes, the stationary lifetime is a reliable measure of the expected node lifetime. For this reason, the stationary lifetime is a useful measure to identify the critical node, which is actually the needed information for our algorithm.

It is not difficult to see that, if node $i$ applies a fixed threshold and all the other nodes have infinite energy supply; the energy consumption at that node, $g_{k,i}$, is stationary. Under this circumstance, the renewal process $\tau_i({\bf e})$ is stationary, and we can apply the elementary renewal theorem \cite{smith1958renewal},
\begin{align}
\lim_{e_i\rightarrow \infty} \frac{T_i(e_i)}{e_i}
   &= \frac{1}{\EE\{g_{k,i}\}}.
\label{Eq.ERTh}
\end{align}
Taking into account that
\begin{align}
\EE \{g_{k,i}\}
    =& \sum_{j=1}^N C_{0,i,j} P\left({a_k}=0| y_k=j\right) p_j
    + \sum_{j=1}^N C_{1,i,j} P\left({a_k}=1| y_k=j\right) p_j \nonumber\\
    =& \sum_{j=1}^N p_j C_{0,i,j} P\left({Q_j} x_k  <  \mu(j) | y_k=j\right)      
    \nonumber\\
    +& \sum_{j=1}^N p_j C_{1,i,j} P\left({Q_j} x_k \ge \mu(j) | y_k=j\right)  \nonumber\\
    =& {{\bf u}_i^\top}{\bf C}_0 {\bf p} 
     + \sum_{j=1}^N p_j (C_{1,i,j}-C_{0,i,j}) F_j\left(\frac{\mu(j)}{{Q_j}}\right)
    \nonumber\\
    =& {{\bf u}_i^\top}({\bf C}_0 + {{\bf \Delta}^\top}{\bf F}) {\bf p}
 \end{align}
where ${\bf u}_i$ is a $N\times1$ unit vector with value one in the $i$-th component and zero in the rest, and diagonal matrix ${\bf F}$ contains diagonal elements 
\begin{equation}
F_j\left(\frac{\mu(j)}{Q_j}\right) = P\left(x_k \ge \frac{\mu(j)}{Q_j} | y_k=j \right).
\label{EqFj}
\end{equation}

Therefore, assuming stationarity, the expected lifetime of node $i$ with battery $e_i$ {(i.e., $T_i(e_i)=\EE\{\tau_i(e_i)\}$)} can be approximated as
\begin{align}
T_i(e_i) \approx \frac{e_i}{\EE\{g_{k,i}\}} 
         = \frac{e_i}{{{\bf u}_i^\top}({\bf C}_0 + {{\bf \Delta}^\top}{\bf F}) {\bf p}}.
\label{EqLifetime}
\end{align}
The last term of the above equation will be useful for the analysis in the next subsections. In practice, given that the importance distribution of messages needs to be known to compute {\bf F} according to \eqref{EqFj}, $T_i(e_i)$ could be estimated by each node in a local and online manner without using \eqref{EqLifetime} just by dividing the available energy by an estimate of the average consumption per time.

Next, a definition for the critical node is proposed, which is useful to characterize the constant-threshold regions. 

\begin{mydefinition}
Given an MDP $({\cal T},{\cal A},{\cal P},r)$ for a network with $N$ nodes and an energy vector ${\bf e}$, $i \in {\cal N}$ is denoted as the \emph{critical node} of the network iff
\begin{align}
i = \arg\min_j T_j(e_j) \quad \forall j \in {\cal N}.
\label{EqNext2Die}
\end{align}
\end{mydefinition}

Note that we use the stationary lifetime instead of the expected lifetime of the nodes as our measure to identify the critical node. However, in a network with constant thresholds the expected lifetime of the critical node is expected to be equal to its stationary lifetime, since the environment is stationary up to the dead of the first node.

Summarizing, we can determine the constant-threshold region corresponding to any energy vector by computing the critical node. The identification of the critical node and the computation of its stationary lifetime is summarized in Algorithm \ref{AlgLife}.

\begin{algorithm}                      
\caption{Compute $(i,T_i) = {\bf CriticalNode}(\bfMu,{\cal G},{\bf e},{\bf p},{\bf C}_0,{\bf C}_1)$} 
\label{AlgLife}                           
\begin{algorithmic}                    
     \STATE ${\bf \Delta} = {\bf C}_1^\top - {\bf C}_0^\top$ 
     \STATE $F_j = P\left(x_k \ge \frac{\mu(j)}{Q_j} | y_k=j \right), \qquad j=1,\cdots,N$
     \STATE $T_j = \dfrac{e_j}{{\bf u}_j^\top({\bf C}_0 + {\bf \Delta}^\top {\bf F}) {\bf p}}, \qquad j=1,\cdots,N$
     \STATE $i = \arg\min_j T_j(e_j)$, \ \ \ $\forall j \in {\cal N}$
\end{algorithmic}
\end{algorithm}


\subsection{{Asymptotic thresholds}}
\label{S:Asymptotic_thresholds} 

In this section we will show how to compute the threshold values in a given region from information about the critical node. To do so, and based on the empirical observations, we will assume that $\lambda( {\bf e} )$ is approximately linear (or, equivalently, $\mu({\bf e},{\bf z})$ does not depend on ${\bf e}$) and determine the coefficients of this linear approximation.

\subsubsection{A linear approximation to $\lambda({\bf e})$}
\label{S:linear_approx}

Initially, we will determine the basic equations coming from the idea that $\lambda({\bf e})$ is approximately linear in each constant region of the energy space, i.e., 
\begin{equation}
\lambda({\bf e}) \approx \hat{\lambda}({\bf e}) = {\bf w}^\top{\bf e} + w_0.
\label{EqLinLam}
\end{equation}
Note that the value of parameters ${\bf w}$ and $w_0$ depends on the specific region of the energy space where the vector ${\bf e}$ is. In order minimize the notation overload, we do not make this dependency explicit.

Using \eqref{EqLinLam} and assumption B1, \eqref{Eq.thresh_theorem} can be rewritten as
\begin{align}
\mu({\bf e},y) \approx {\mu(y)=}  {\bf w}^\top(\EE\{{\bf c}_1|y\} - \EE\{{\bf c}_0|y\}) 
\label{Eq.Mu_Const}
\end{align}
and \eqref{Eq.Th1lambda} as
\begin{align}
{{\bf w}^\top\EE\{{\bf c}_0\} = \EE\{(Q_{y} x-\mu(y))^+ \}.}
\label{Eq.Lambda_Lin}
\end{align}

Using vector notation, if $\bfMu$ is a vector with components {$\mu(y)$}, we can write \eqref{Eq.Mu_Const} and \eqref{Eq.Lambda_Lin} as
\begin{align}
\bfMu &= {{\bf \Delta}\cdot}{\bf w}
\label{Eq.Mu_Const2}
\\
{\bf p}^\top {\bf C}_0^\top {\bf w} &= {\bf p}^\top {\bf h}(\bfMu),
\label{Eq.Lambda_Lin2}
\end{align}
where ${\bf h}(\bfMu)$ is a vector with components
\begin{align}
{h_j(\mu(j)) = \EE\{(Q_j x-\mu(j))^+ |y=j \}.}
\label{Eq.DefHi}
\end{align}

\subsubsection{Computing the asymptotic thresholds}
\label{SecCompMu}

The asymptotic thresholds $\bfMu$ can be computed using \eqref{Eq.Mu_Const2} for a given slope vector ${\bf w}$, which {according to \eqref{Eq.Lambda_Lin2}} must be a solution of
\begin{align}
{\bf p}^\top {\bf C}_0^\top {\bf w} = {\bf p}^\top {\bf h}({\bf \Delta}\cdot{\bf w}).
\label{Eq.Lambda_Lin3}
\end{align}

Note that $w_0$ in \eqref{EqLinLam} is not required to compute the thresholds since $\bfMu$ is the difference of two shifted versions of $\lambda({\bf e})$. Hence, computing $\bfMu$ converts into computing ${\bf w}$. Although \eqref{Eq.Lambda_Lin3} does not have a unique solution, we will show that the knowledge of the critical nodes (more specifically, the sequence of critical nodes up to the network dead) determines ${\bf w}$ uniquely. The pseudocode of the algorithm \textbf{Thresholds}, which computes ${\bf w}$ and $\bfMu$ for a given energy vector, is shown in Algorithm \ref{alg1} and its explanation is the main goal of this section.

\begin{algorithm}                      
\caption{Compute 
         {$(\bfMu,{\bf w}) = {\bf Thresholds}(i,T_i,{\cal G},{\bf e},{\bf p},{\bf C}_0,{\bf C}_1)$}}    
\label{alg1}                           
\begin{algorithmic}                    
    
    \STATE $(\calD,\calS) = \text{split}_{\cal G}(i)$
    \STATE ${\bf \Delta} = {\bf C}_1^\top - {\bf C}_0^\top$
    \STATE $\overline{\bf c} = {\bf C}_0 {\bf p}$
    \newline \newline        
    \COMMENT{Step 1: Compute ${\bf w}_\calS$}
    \IF{$\calS=\emptyset$}             
        \STATE ${\bf w}_\calS=\emptyset$
        \STATE $\alpha = 0$
        \STATE $\boldsymbol{\beta} = {\bf 0}$
    \ELSE
        \STATE $\tilde{\bf e}_\calS 
                    = {\bf e}_S 
                    - T_i ({\bf C}_{0\calS\dot\calS}
                          +({\bf C}_{0\calS\dot\calS}+{\bf C}_{1\calS\dot\calS}){\bf F}_{\dot\calS})
                           {\bf p}_{\dot\calS}$
        \STATE $\tilde{p}_0 = 1 - {\bf 1}^\top {\bf p}_\calS$
        \STATE $\tilde{\bf p}_{\dot\calS} = (\tilde{p}_0, {\bf p}_\calS^\top)^\top$
		\STATE $(\tilde{\bfMu}_S,{\bf w}_{\cal S})={\bf main}({\cal G}_\calS,\tilde{\bf e}_\calS,\tilde{\bf p}_{\dot\calS},{\bf C}_{0\calS\dot\calS},{\bf C}_{1\calS\dot\calS})$

        \STATE $\alpha = \overline{\bf c}_\calS^\top {\bf w}_\calS $
        \STATE $\boldsymbol{\beta} = {\bf \Delta}_\calS {\bf w}_\calS$
    \ENDIF     
    \newline\newline    
    \COMMENT{Step 2: Compute $w_i$}
    \STATE Solve $\overline{c}_i w_i + \alpha = {\bf p}^\top {\bf h}({\bf \Delta}_i w_i + \boldsymbol{\beta})$ for $w_i$ .
    \newline\newline   
    \COMMENT{Step 3: Compute ${\bf w}_\calD$}
    \IF{$\calD=\emptyset$}                   
        \STATE ${\bf w}_\calD = \emptyset$
    \ELSE 
        \STATE ${\bf w}_\calD = {\bf 0}$ 
    \ENDIF    
    \newline
    \STATE Compose ${\bf w}$ from its components  \{$w_i$, ${\bf w}_\calD$ and ${\bf w}_\calS$\}
    \STATE {$\bfMu = {\bf \Delta}\cdot{\bf w}$}

\end{algorithmic}
\end{algorithm}


The key idea of the algorithm \textbf{Thresholds} is to split the set of the network nodes $\mathcal{N}$ into three different subsets: the \emph{critical node} $i$ which is previously computed in the call to the function \textbf{CriticalNode}, the set of \emph{disconnected nodes,} $\calD$ (i.e., the nodes that would get disconnected from the sink when the critical node died), and the set of \emph{surviving nodes} $\calS$ (i.e., the nodes that would keep connected to the sink after the battery depletion of the critical node). This is represented in the pseudocode by function ${\bf split}_{\cal G}(i)$ in Algorithm \ref{alg1}, which makes use of the network graph ${\cal G}$. From this partition we will compute the different components of $\mathbf{w}$ as explained below.


Using the subindex notation defined in Section \ref{SecNotation}, the slope vector ${\bf w}$ can be subsequently split into three components linked to the different node subsets:  $w_i$, ${\bf w}_\calD$ and ${\bf w}_\calS$.

Thus, \eqref{Eq.Mu_Const2} can be written as

\begin{equation}
\bfMu= {\bf \Delta}_{\calS} {\bf w}_\calS + {\bf \Delta}_{\calD} {\bf w}_\calD +{\Delta}_{i} w_i.
\label{w_composite}
\end{equation}


According to \eqref{EqLinLam}, the total expected reward must grow linearly with the energy available at any node. However, if $i$ is the critical node, the contribution of any node $j\in \calD$ to the total reward is constrained by the energy at node $i$. An increment of the battery charge at node $j$ does not influence the total reward, since the extra energy cannot be used after the death of the critical node $i$. Mathematically, this implies that

\begin{equation}
{\bf w}_\calD = 0.
\label{wj_zero}
\end{equation}

Thus, taking into account \eqref{wj_zero}, and defining the expected cost vector $\overline{\bf c}$ as
\begin{align}
\overline{\bf c} = {\bf C}_0 {\bf p},
\label{MeanC0}
\end{align}
where ${\bf p}$ is a vector with components $p_j=P\left(y=j\right)$, \eqref{Eq.Lambda_Lin3} can be rewritten as
\begin{align}
\overline{c}_i w_i + \overline{\bf c}_\calS^\top {\bf w}_\calS 
   = {\bf p}^\top {\bf h}({\bf \Delta}_i w_i + {\bf \Delta}_\calS {\bf w}_\calS).
\label{EqB2}
\end{align}

For a given ${\bf w}_\calS$, \eqref{EqB2} can be solved for finding ${\bf w}_i$. Moreover, the solution is unique. This can be shown by noting that
\begin{enumerate}
\item Since $\overline{c}_i>0$, the left-hand side of \eqref{EqB2} is a growing function of $w_i$.
\item Since the right-hand side of \eqref{EqB2} can be written as
\begin{equation}
{\bf p}^\top{\bf h}({\bf \Delta}_\calS {\bf w}_\calS + {\bf \Delta}_i w_i) =
      \sum_{j=1}^{N} p_j h_j({\bf \Delta}_{\calS j} {\bf w}_\calS + {\Delta}_{ij} w_i),
\end{equation}
where $h_j$ are non-increasing functions and ${\Delta}_{ij}>=0$ and ${\Delta}_{ii}>0$, the right-hand side is a non-increasing function.
\end{enumerate}

Up to this point, only the slope for the surviving nodes, ${\bf w}_\calS$, is left to be computed. To do so, note that, if $e_j\gg e_i$ $\forall j\neq i$, node $i$ will contribute to the total reward for a small fraction of time, and, thus, the relative influence of node $i$ (and of all nodes $j \in \calD$) on the total reward is negligible. Mathematically, this implies that we can compute the weights ${\bf w}_{\cal S}$ by assuming $e_i=0$ (and, consequently, that all nodes $j \in \calD$ are disconnected from the network). As a consequence, computing ${\bf w}_{\cal S}$ becomes equivalent to computing the weights of the subnetwork containing only the nodes in $\calS$, by recursively calling the threshold estimation algorithm, as shown in Algorithm \ref{alg1}.

Note that, if the assumption $e_j \gg e_i$ does not hold, the influence of the critical node in ${\bf w}_\calS$ is not negligible, and the estimation of ${\bf w}_\calS$ (and $\bfMu$) becomes inaccurate. In the example of Fig. \ref{fig:MuExp}, this happens around $e_1\approx e_2$, i.e., in the transition region where the threshold values are different from those of the constant regions. However, the proposed estimation procedure provides a good threshold estimate in the constant regions, which covers the majority of the energy space.

\subsubsection{Recursive computation}

Algorithm \ref{alg1} uses a recursive call to function \textbf{main} in the form
\begin{equation}
 (\cdot,{\bf w}_{\cal S})={\bf main}({\cal G}_\calS,\tilde{\bf e}_\calS,\tilde{\bf p}_{\dot\calS},
                          \tilde{\bf C}_{0\calS\dot\calS},\tilde{\bf C}_{1\calS\dot\calS}).
                        \label{Eq.wSfS}
\end{equation}

Note that, in this recursion, we must proceed with caution because the arguments of this call are not a simple restriction of {its arguments to the subgraph containing the surviving nodes}. The parameters of function \textbf{main} should represent the expected state and the behavior of the network \emph{after} the dead of node $i$. More specifically, ${\cal G}_\calS$ is the graph of the surviving network, and $\dot{\calS}=\calS \cup \{0\}$ (i.e., the dot is used to remark {that the information about the event of no messages in the network is included}), $\tilde{\bf e}_\calS$ is the expected energy vector after dying the critical node, and 
$\tilde{\bf p}_{\dot\calS}$, $\tilde{\bf C}_{0\calS\dot\calS}$ and $\tilde{\bf C}_{1\calS\dot\calS}$ should reflect the source probabilities and the expected costs after that time.

Using assumptions B1 and B3, we can take $\tilde{\bf C}_{0\calS\dot\calS} = {\bf C}_{0\calS\dot\calS}$ and $\tilde{\bf C}_{1\calS\dot\calS} = {\bf C}_{1\calS\dot\calS}$ because there are not route changes in the surviving network graph. 

The energy vector $\tilde{\bf e}_{\calS}$ must discount the expected consumption up to the battery depletion of node $i$. From \eqref{EqLifetime}, we can define the expected consumption of the surviving nodes as
\begin{equation}
\bar{\bf g}_{\calS} = (\tilde{\bf C}_{0\calS\calS} + \tilde{\bf \Delta}_{\calS\calS}^\top{\bf F}_\calS) \tilde{\bf p}_\calS,
\end{equation}
and, considering the expected node lifetime of node $i$, $T_i$, the expected amount of available energy at the surviving nodes when node $i$ dies can be computed as
\begin{equation}
\tilde{\bf e}_\calS = {\bf e}_\calS - T_i \bar{\bf g}_{\calS}.
\end{equation}

The frequency of message generation at each node keeps unchanged (and, thus $\tilde{\bf p}_{\dot\calS}$ in \eqref{Eq.wSfS} is equal to ${\bf p}_{\dot\calS}$ in the initial call, except for $p_0$, which must also integrate the probability of message generation in any node of the disconnected set). There are intervals with no messages in the subnetwork because some nodes are disconnected.

\begin{figure}[ht]
\centering
\includegraphics[clip,width=1\columnwidth]{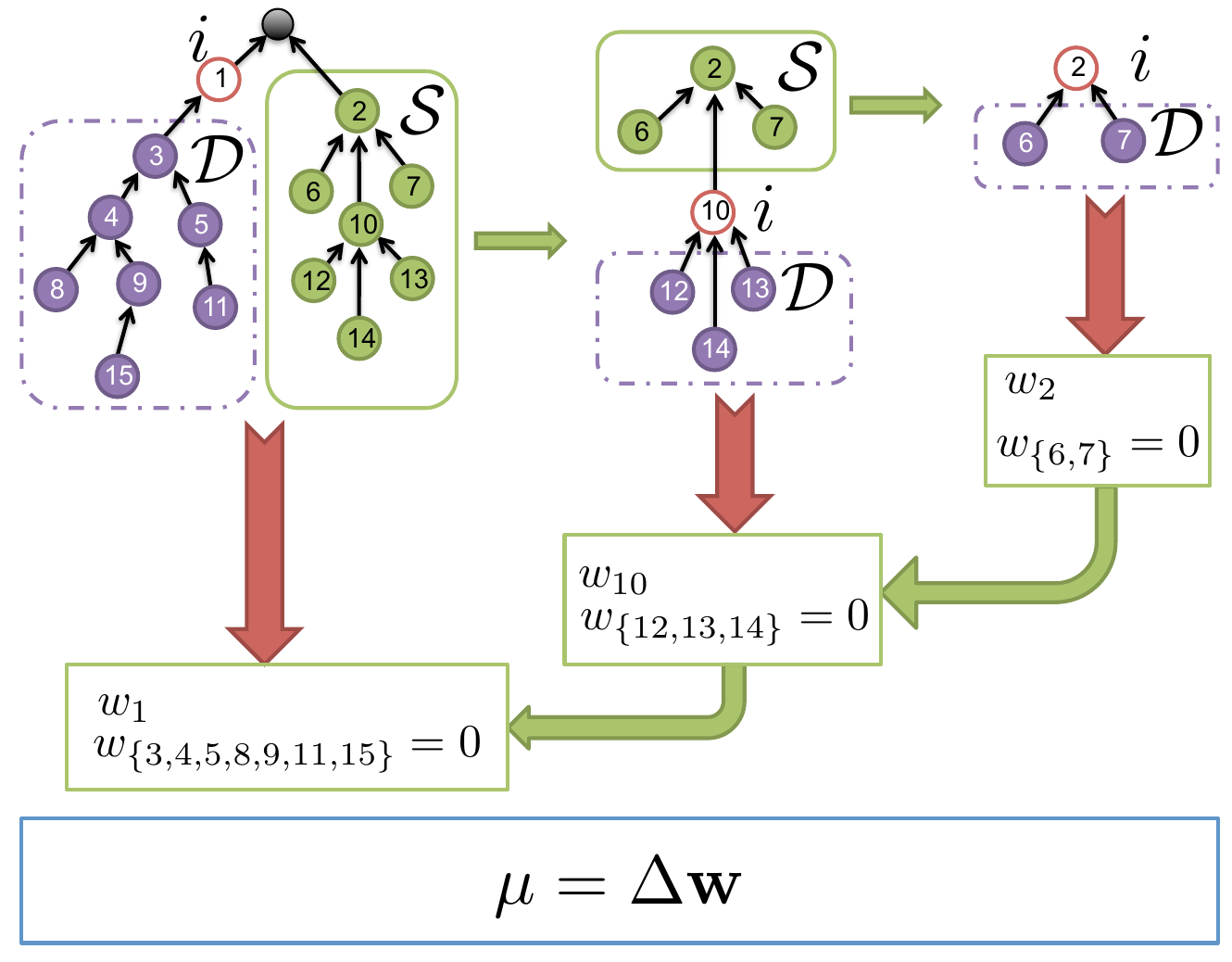}
\caption{Sketch for the computation of the node thresholds.}
\label{fig:AlgThresholds}
\end{figure}

The recursive estimation process in the \textbf{main} algorithm is illustrated in Fig. \ref{fig:AlgThresholds}. {Initially, node $1$ is identified as the critical node (the first node to die), so $w_1$ and ${\bf w}_{\calD}$ with $\calD=\{3,4,5,8,9,11,15\}$ can be computed. Then, we apply the same procedure to the subsequent surviving nodes until there is no surviving nodes to identify (i.e., $\calS=\emptyset$).} {Note that in order to compute $w_1$ we need to know ${\bf w}_\calS$ with $\calS=\{2,6,7,10,12,13,14\}$. This involves first the computation of $w_2$, and subsequently the computation of $w_{10}$ so that we can later compute $w_1$, i.e., a forward computation is performed.}   

Finally, it is worth mentioning that assumption B3 can be relaxed. The restriction to tree-topology (or fixed-route) networks has been used in the derivation of parameters $\tilde{\bf p}_{\dot\calS}$, $\tilde{\bf C}_{0\calS\dot\calS}$ and $\tilde{\bf C}_{1\calS\dot\calS}$ in the recursive call to function \textbf{main}. But the proposed algorithm could be extended to general arbitrary topologies with dynamic routing, provided that we can compute the expected cost matrices and source probabilities for those networks. Our choice of the tree topology has been motivated by simplicity reasons. In that case assumption B1 should also be relaxed to allow costs depend on the network topology, and not only on the message source.



\section{Experimental evaluation}
\label{S:Simulations}

In this section we present some simulation results to show the performance of the proposed algorithm. 
Hereunder, some common features of the experimental setup are exposed.

\begin{enumerate}
\item
The energy consumption of the nodes, which is associated to their operational modes, are: (a) $E_S$, energy spent on sensing a message; (b) $E_R$, energy spent on receiving a message; and (c) $E_T$, energy spent on transmitting a message. Unless otherwise stated, we use a deterministic energy model given by constant and known parameters: Energy values are set to $E_T=5$, $E_R=5$ and $E_S=1$. Matrices ${\bf C}_0$ and ${\bf C}_1$ are built according to the transmission routes as follows: ${\bf C}_0$=$E_S\bf I$, where $\bf I$ is an $N\times N$ identity matrix; and ${\bf C}_1$=${\bf C}_0$+$E_R \textbf{T} (\textbf{1}-\textbf{I}) + E_T \textbf{T}$, where $\bf T$ is the $N\times N$ routing matrix with components $t_{i,j}=1$ if a message generated by node $j$ is routed to the sink through node $i$ and it is $0$ otherwise, and $\textbf{1}$ is a $N\times N$ all-ones matrix.

\item Nodes are homogeneous and their initial battery level is the same (and equal to $10000$) except for the sink, which has unlimited power supply. Nodes keep working until their batteries expire. The network dies when all the sink neighbors have died.

\item Source nodes are selected at random with equal probability. Further, messages are generated according to an exponential distribution with mean $1$, independently of the source node. Moreover, when all the nodes are still alive, $P(y_k=0)=0$~.

\item The behavior of the proposed algorithm, named \textit{Global Cooperative Transmitter} (GCT), is compared with other types of censoring (selective) strategies. In particular, we consider the following ones: (a) \textit{Non-Selective} (NS) strategy, in which the sensor nodes do not censor any generated message; (b) \textit{Selective Transmitter} (ST) strategy, where nodes only consider in the decision-making process their own local information (i.e., the energy consumed during the different operational modes, the available battery, and the message importance) \cite{ArroyoVallesEtAl09}; (c) \textit{Selective Forwarder} (SF) strategy, where nodes also take into account the one-hop neighbor's behavior regarding the cooperation in the retransmission of messages \cite{ArroyoVallesEtAl11}; and (d) the \textit{Cooperative Transmitter} (CT) strategy, which is mainly designed for line networks, and assumes that the first node that dies is the one connected to the sink \cite{Fernandez-Bes11}.

\item Performance is assessed in terms of the importance sum of all messages received by the sink, the number of receptions at the sink, and the number of \emph{generated} messages (the latter amounts to measure the network lifetime).

\item Experimental results are averaged over 100 different simulation runs.
\end{enumerate}

\subsection {Line network}
\begin{figure}[ht]
\centering
\includegraphics[clip,width=1\columnwidth]{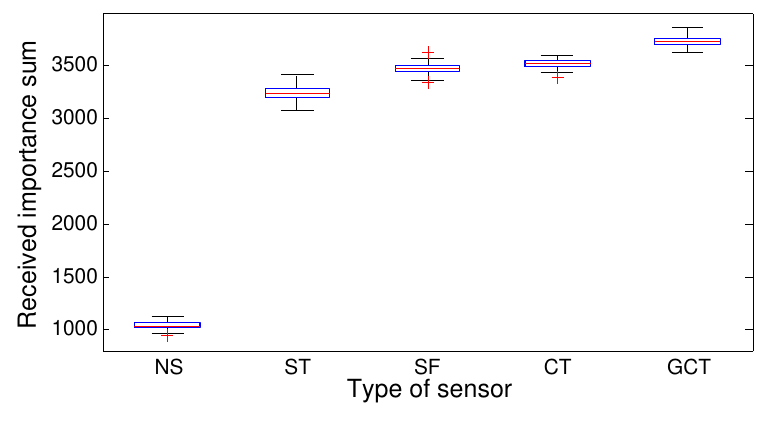}
\caption[Line10nodes]{Received importance sum at the sink in a line network composed of 10 nodes.}
\label{fig:Line10nodes}
\end{figure}

In the first setup, 10 nodes are placed in a line. Each node only communicates with the two adjoint neighbors, and the last node is connected to the sink. Fig. \ref{fig:Line10nodes} shows the performance of the different strategies in terms of the received importance sum at the sink. The better performance of the GCT algorithm can be noticed. As expected from previous results \cite{Fernandez-Bes11}, the SF algorithm outperforms the ST one, and both, the ST and the SF algorithms, are outperformed by the CT algorithm. However, the improvement of the CT algorithm regarding the SF one is not so striking. Further, and looking at Table \ref{Table.Line10nodes}, the network lifetime of the GCT algorithm is longer than in the other approaches because the number of generated messages is larger. Clearly, the worst performance corresponds to the NS algorithm since it does not discard any message. The network lifetime shows the same trend that the one observed for the received importance sum for the remaining selective algorithms: the CT algorithm achieves a longer lifetime than the SF one, and in turn, the latter lasts longer than the ST algorithm. There is a quantitative gap between the ST and SF algorithms, where only local and first-hop neighboring information is available, respectively, and the CT and GCT algorithms, where more information is shared among nodes. Hence, the more information is available at the nodes, the better the performance is.
 
\begin{figure}[ht]
\centering
\includegraphics[clip,width=1\columnwidth]{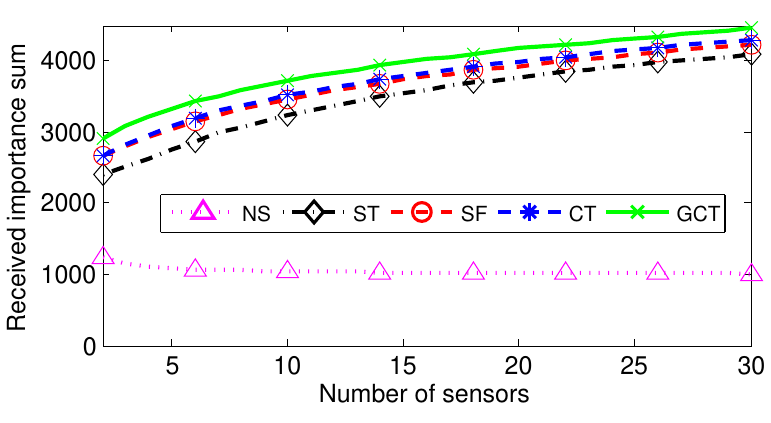}
\caption[Line10nDiffNumSensors]{Received importance sum at the sink in a line network topology composed of different number of nodes.}
\label{fig:LineDiffNumbernodes}
\end{figure}

\begin{figure}[ht]
\centering
\includegraphics[clip,width=1\columnwidth]{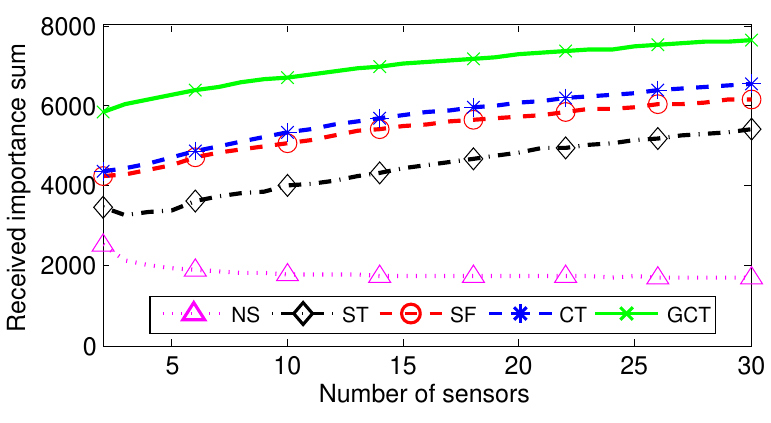}
\caption[Line10nodesLowET]{Received importance sum at the sink in a line network topology composed of different number of nodes for a low transmission energy consumption.}
\label{fig:LineDiffNumSensorsLowET}
\end{figure}

Fig. \ref{fig:LineDiffNumbernodes} illustrates the received importance sum at the sink for a line network composed of different number of nodes. The larger the line-network size is, the higher the received importance sum is for all the selective algorithms.
Once again, the GCT algorithm outperforms the others. Notice that the performance achieved by the CT algorithm is slightly better than the one achieved by the SF algorithm, and the later outperforms the ST one. 
Fig. \ref{fig:LineDiffNumSensorsLowET} shows the performance results for the previous setup but decreasing the energy consumption due to a transmission ($E_T=1$). Now, differences between the GCT algorithm and the remaining selective ones (i.e., CT, SF and ST) are much more significant. Also, notice that the received importance sum is even larger than before as the cost of transmitting reduces. 

\begin{figure}[ht]
\centering
\includegraphics[clip,width=1\columnwidth]{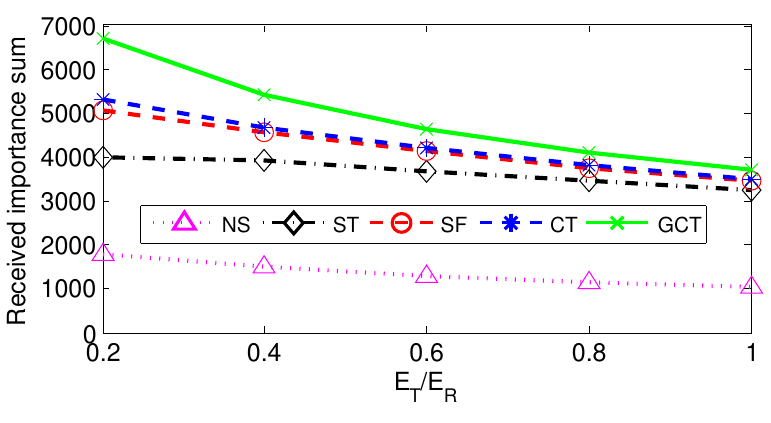}
\caption[Line10nodesRatioETER]{Received importance sum at the sink in a 10-node line network, when $E_T$ varies from $1$ to $5$ and $E_R=5$.}
\label{fig:Line10nodesRatioETER}
\end{figure}

According to the previous results, the performance depends on the values of $E_T$ and $E_R$. For this reason, next we analyze the ratio $E_T/E_R$ on a 10-node line network, when $E_R=5$ and $E_T$ varies from $1$ to $5$. Some sensors, mainly low-power ones, require low consumption in the transmissions. The consumption values may be even lower than those due to receptions \cite{Kyaw08, Brownfield10}. Fig. \ref{fig:Line10nodesRatioETER} reveals that the performance gain achieved by the GCT reduces as the ratio $E_T/E_R$ increases (i.e., both costs tend to be similar). 

\subsection {Tree network}
\begin{figure}[ht]
\centering
\includegraphics[clip,width=1\columnwidth]{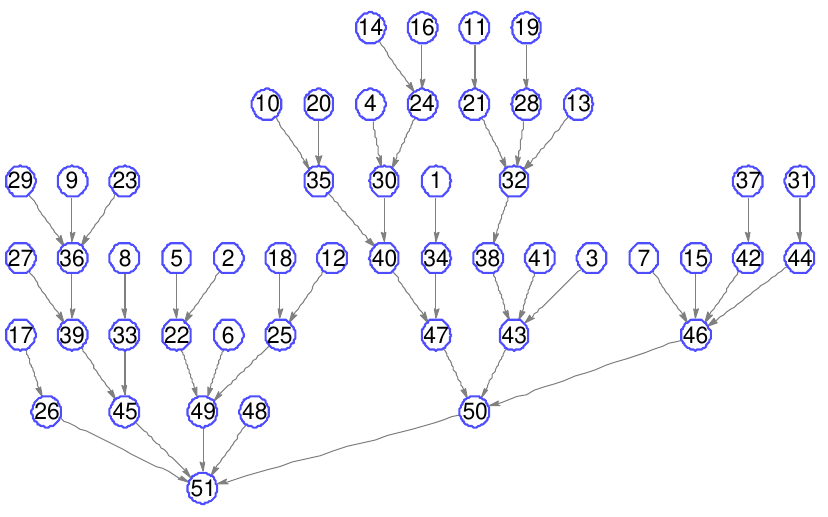}
\caption[TreeTopology]{Example of a tree network topology.}
\label{fig:TreeTopology}
\end{figure}

In the second setup, a tree-network topology where 50 nodes are randomly deployed is considered. The network is built as follows: each node of index $i$ chooses randomly with equal probability another node of index $j$ (so that $j>i$) to be connected. Fig. \ref{fig:TreeTopology} shows an example of a random network topology, where node $51$ is the sink (the root).    

\begin{figure}[tt]
\centering
\includegraphics[clip,width=1\columnwidth]{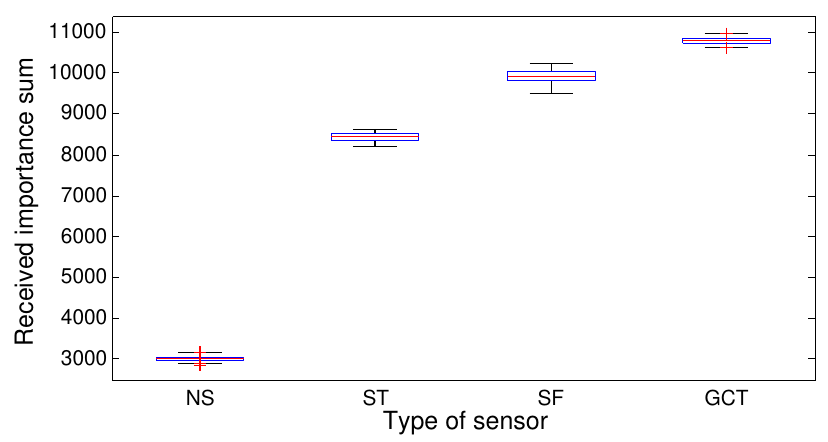}
\caption[TreeImpSum]{Received importance sum at the sink in a tree network topology composed of 50 nodes where the transmission energy consumption is different at every node.}
\label{fig:TreeImpSum}
\end{figure}

Fig. \ref{fig:TreeImpSum} depicts the received importance sum at the sink in the tree-network scenario for the different algorithms. Results are averaged over 100 different tree-network topologies. We do not include the results corresponding to the CT algorithm since it was mainly designed for line networks (see \cite{Fernandez-Bes11} for further details). Besides, the energy consumption due to transmissions varies from one node to another. $E_T$ values are drawn from a uniform distribution between $5$ and $20$. There are several reasons for a node to consume more energy in a transmission, such as the need to transmit at a higher power or the higher number of retransmissions that it has to carry out to successfully send the message, to name a few. From the plot, it can be seen that the best performance corresponds to the GCT algorithm, which shows a significant gain with regard to the other selective networks. 

\section{Conclusions}\label{S:Conclusions}

In this paper we {analyzed} the problem of cooperative policies for energy efficiency in sensor networks. The formulation of the problem using Markov Decision Processes demonstrated that, although the computation of the optimal policies stated severe computational problems, for large values of the initial battery load the optimal policy can be approximated by a finite collection of constant thresholds. Our experiments showed that cooperative policies based on the estimation of these thresholds obtained a noticeable performance gain with respect to previous state-of-the-art schemes.

However, the application of the censoring algorithm in a practical scenario presents some major difficulties:
\begin{itemize}
\item The recursive computation of the asymptotic thresholds requires to estimate the expected evolution of the network graph. Not only is the critical node required, but also the sequence of critical nodes of the successive surviving subnetworks, up to the end of the network lifetime.
\item The algorithm is not decentralized: the asymptotic thresholds must be computed at a central node gathering all the information from all the nodes.
\item The algorithm is not adaptive: it requires some information, e.g. function ${\bf h}$, that may not be available at any node.
\end{itemize}

Thus, the design of decentralized, adaptive and computationally efficient cooperative censoring policies that allow dynamic routing is a major challenge that remains to be solved.

\section*{Appendix A: Proof of Theorem \ref{Th.General}}
\label{ProofThGeneral}

Using \eqref{EqReward} in the Bellman equation (see \eqref{Eq.Bellman}), and noting that ${\bf s}=({\bf e},{\bf z})$ we get
{\begin{align}
V_{\pi^{*}}({\bf s}) 
   &= \max_{a \in \{0,1\}} 
           \left\{ a x \EE\left\{q_k | a_k=a, {\bf e}_k={\bf e}, {\bf z}_k={\bf z} \right\} \right.
           \nonumber \\
    +&     \left. \EE\left\{ 
                  V_{\pi^{*}}({\bf e}_{k+1},{\bf z}_{k+1})|a_k = a, {\bf e}_k={\bf e}, {\bf z}_k={\bf z} 
              \right\}  \right\}  \nonumber\\
   =& \max_{a \in \{0,1\}} 
           \{ a x Q({\bf e},{\bf z})  
    +     a  \EE\left\{ V_{\pi^{*}}(({\bf e}-{\bf c}_{1k})^+,{\bf z}_{k+1}) | {\bf z}_k={\bf z} \right\}  
           \nonumber\\
    +& (1-a) \EE\left\{V_{\pi^{*}}(({\bf e}-{\bf c}_{0k})^+,{\bf z}_{k+1}) | {\bf z}_k={\bf z} \right\}
            \}.  
\label{Eq.Bellman2}
\end{align}}

Defining $\lambda({\bf e})$ as in \eqref{EqDefLambda}, and taking into account that ${\bf z}_k$ is a stationary and time independent sequence {and that ${\bf c}_{0k}$ and ${\bf c}_{1k}$ are stochastic stationary and i.i.d. processes},
{\begin{align}
V_{\pi^{*}}({\bf s}) 
   = \max_{a \in \{0,1\}} &
          \{ a x Q({\bf e},{\bf z})  
    +     a     \EE\left\{ \lambda(({\bf e}-{\bf c}_{1k})^+) | {\bf z}_k={\bf z} \right\}  
           \nonumber\\
   &+ (1-a) \EE\left\{\lambda(({\bf e}-{\bf c}_{0k})^+) | {\bf z}_k={\bf z} \right\}
           \}. 
\label{Eq.Bellman3}
\end{align}}
Defining the threshold function as in \eqref{Eq.thresh_theorem}, we get
{\begin{align}
V_{\pi^{*}}({\bf s}) 
   =  \EE\left\{ \lambda(({\bf e}-{\bf c}_{0k})^+) | {\bf z}_k={\bf z} \right\} 
    + \max_{a \in \{0,1\}} 
          \{a \left(x Q({\bf e},{\bf z}) - \mu({\bf e},{\bf z}) \right) 
           \}. 
\label{Eq.Bellman4}
\end{align}}
Thus, the maximum in \eqref{Eq.Bellman4} is achieved for action $a$ given by \eqref{Eq.opt.dk}. So we can write
{\begin{align}
V_{\pi^{*}}({\bf s}) 
   = \EE\left\{ \lambda(({\bf e}-{\bf c}_{0k})^+) | {\bf z}_k={\bf z} \right\} 
   + \left(x Q({\bf s}) - \mu({\bf s})\right)^+.  
\label{Eq.Bellman5}
\end{align}}
Finally, taking the expectation of $V_{\pi^{*}}({\bf e},{\bf z}_{0})$ with respect to ${\bf z}_{0}$, we get \eqref{Eq.Th1lambda}.

\section*{Acknowledgments}			
The work in this paper was supported by Spanish Government Grant no. TEC2011-22480 and PRI-PIBIN-2011-1266. The work of Fernandez-Bes was also supported by the Spanish MECD FPU program. The work of Arroyo-Valles was also supported by the Spanish MECD under the National Program of Human Resources Mobility from the I+D+i 2008-2011 National plan. 





\bibliographystyle{model1-num-names}
\bibliography{Bibliography}








\newpage
\begin{table}[t]
\begin{center}
\renewcommand\baselinestretch{1}
\caption{Summary of the most significant notation.}
\label{Table_Notation}
\renewcommand\baselinestretch{1.5}
\fontsize{8}{8}\selectfont
\vspace{0.3cm}
\begin{tabular}{|l|l|}\hline 
Symbol & Meaning \\ \hline \hline
$k$ & Time epoch index\\ 
${\bf e}_k$, {$e_{k,i}$} & Energy vector at time epoch $k$/ energy at time epoch $k$ and sensor $i$\\
$y_k$& Index of the message source at epoch $k$\\
${\bf z}_k$ &  Vector containing {(at least) the message importance and the source node at time $k$}\\
${\bf s}_k$ & State vector at epoch $k$\\
$a_k$ & Decision at time $k$\\
$\pi_k$ & Transmission policy at time $k$\\
${\bf c}_{1k}$,${\bf c}_{0k}$        & {Energy consumed by nodes at epoch $k$ when the network transmits/censors a message} \\
${\bf c}_{1k}(j)$, ${\bf c}_{0k}(j)$ & {Idem when the source node is $j$}\\
{$Q({\bf s})$, $Q_y$} & {Probability of success for any state ${\bf s}$ / Probability of success when the source is $y$} \\
$p_j$, ${\bf p}$ & Probability that node $j$ is the source/ probability vector of being a source\\
$\mu({\bf s})$, $\bfMu$ & Decision threshold/ decision threshold vector \\
$\lambda({\bf e})$ & Total reward that can be expected for the initial energy vector ${\bf e}$\\
$T_i(e_i)$ & Stationary lifetime of node $i$ for the initial energy $e_i$\\
${\bf M}_\mathcal{J}$ & Submatrix {of ${\bf M}$ taking the columns corresponding to indices in $\mathcal{J}$}\\
${\bf M}_{\mathcal{IJ}}$ & Submatrix {of ${\bf M}$ taking the rows in $\mathcal{I}$ and the columns in $\mathcal{J}$}\\
${\bf C}_1$, ${\bf C}_0$ & Energy consumption matrix for transmitting/censoring a message\\
${\bf \Delta}$ & Matrix containing the incremental costs of deciding to transmit\\
${\cal G}$ & Network graph\\
${\bf w}$ & slope vector for the linear approximation of $\lambda({\bf e})$\\
$\cal D$,$\cal S$ & Subset of disconnected/surviving nodes when the critical node dies\\
$\dot{\calS}$ & Subset of surviving nodes $\cup$ no event in the network\\
\hline
\end{tabular}
\end{center}
\end{table}

\newpage
\begin {table*}[ht]
\fontsize{10}{10}\selectfont
\centering
\begin{tabular}{|c|c|c|c|}
\hline
\hline
 \noalign{\smallskip} \raisebox{0ex}[0cm][0cm]{Algorithm}  & \multicolumn{1}{|c|}{Generated messages} & \multicolumn{1}{|c|}{Received messages} & \multicolumn{1}{|c|}{Discarded messages}\\
\hline
\hline NS& $1042.70$ & $1041.70$ & $0$\\
\hline ST& $13245.40$ & $876.13$ & $12350.23$\\
\hline SF& $17524.58$ & $869.19$ & $16654.39$\\
\hline CT& $23069.05$ & $809.98$ & $22258.07$\\
\hline GCT& $25250.22$ & $943.99$ & $24305.23$\\
\hline
\end{tabular}
\caption{Average number of generated, received (by the sink) and discarded messages for the 10-nodes line network.}
\label{Table.Line10nodes}
\end{table*}

\end{document}